# System-in-the-loop Design Space Exploration for Efficient Communication in Large-scale IoT-based Warehouse Systems

**Robert Falkenberg, Jens Drenhaus, Benjamin Sliwa and Christian Wietfeld**
Communication Networks Institute
TU Dortmund University, 44227 Dortmund, Germany
Email: {Robert.Falkenberg, Jens.Drenhaus, Benjamin.Sliwa, Christian.Wietfeld}@tu-dortmund.de

***Abstract*—Instead of treating inventory items as static resources, future intelligent warehouses will transcend containers to Cyber Physical Systems (CPS) that actively and autonomously participate in the optimization of the logistical processes. Consequently, new challenges that are system-immanent for the massive Internet of Things (IoT) context, such as channel access in a shared communication medium, have to be addressed. In this paper, we present a multi-methodological system model that brings together testbed experiments for measuring real hardware properties and simulative evaluations for large-scale considerations. As an example case study, we will particularly focus on parametrization of the 802.15.4-based radio communication system, which has to be energy-efficient due to scarce amount of harvested energy, but avoid latencies for the maintenance of scalability of the overlaying warehouse system. The results show, that a modification of the initial backoff time can lead to both, energy and time savings in the order of 50% compared to the standard.**

## I. Introduction and Related Work

This paper will contribute to system design and system optimization of IoT-based warehouses and materials handling systems, which embraces a physical testbed PhyNetLab [1] into a simulative design space exploration. The research is part of the project A4 ”Resource efficient and distributed platforms for integrative data analysis” in the collaborative research center SFB 876 ”Providing information by resource-constrained data analysis”, which is sponsored by Deutsche Forschungsgemeinschaft (DFG).

A promising application field for distributed connected platforms are future logistics processes, which waive any central controlling units like inventory management. Instead, such warehouses will be composed of numerous communicating Cyber Physical Systems (CPS), e.g., robots or smart containers with knowledge of their tasks or contained goods and which independently and autonomously reconcile each other to perform their duties. A real-life system testbed for such autonomous materials handling and warehousing is PhyNetLab (cf. Fig. 1), which consists of a large number of containers, each attached with an embedded system platform – the PhyNode [2]. The main components of the PhyNode are a small microcontroller, a photovoltaic cell, a rechargeable battery, and a low-energy transceiver which operates in the 868 MHz Short Range Devices (SRD) band. The testbed enables researchers to address manifold challenges in fields of materials handling, system-, software-, and hardware design, energy management and efficient communication, and evaluate these approaches on a physical large-scale system. Conversely, knowledge from trail runs and actual hardware behavior like sensor data, communication statistics, and energy consumption, is feed back into simulations to refine the system models an increase the simulation accuracy.

In this paper we will present a system-in-the loop simulation model for a typical logistics use case, which incorporates actual hardware characteristics of the PhyNetLab testbed and analyze the scalability and energy-efficiency of the wireless communication system.

### A. Challenges

The overlaying challenge of IoT-based systems like PhyNetLab is an efficient management the available resources, e.g., memory, energy, or radio resources, which are typically very scarce on embedded systems. Since the components of such systems are mobile and not connected to any power grid, the energy management becomes one of the most important design parameters, which in turn constraints the activity of other system components. Although a limited number of mobile robots might autonomously seek for a charging station in case of a low battery state, a recharge-management for thousands of smart containers might render the system unaffordable. As an alternative such containers might harvest the required energy from the environment, such as light. However, the amount of harvested energy in a dimly lit warehouse still requires an extremely efficient and astute system platform. One of the largest energy consumers in such distributed platforms is the communication module, which is typically a radio transceiver.

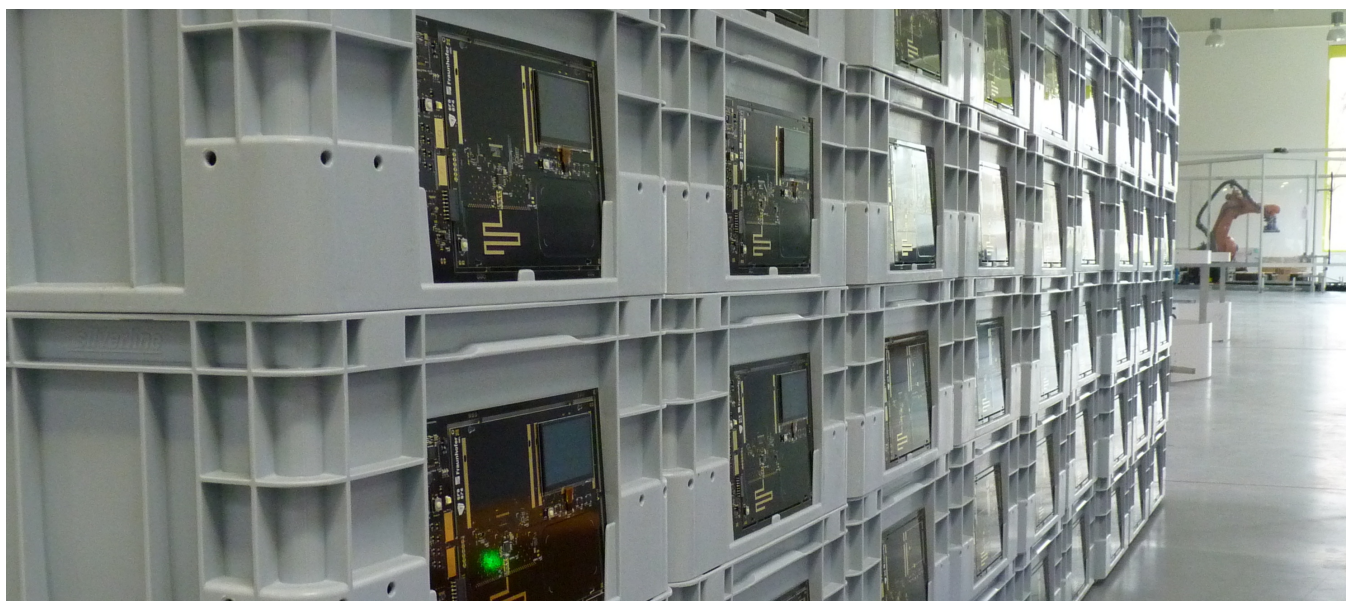

Fig. 1. Photograph of the PhyNetLab, an IoT-based warehouse testbed, which consists of a large-scale deployment of energy-harvesting and communicating smart containers (PhyNodes).

Our recent empirical study [3] points out a quickly and significantly increasing power draw for the standardized assessment of a clear radio channel, if the number of concurrently accessing participants is raised beyond a few dozen nodes due to congestions. From this result arises the demand for an optimized radio channel access scheme, which is scalable in the order of hundreds or thousands of nodes without overtaxing their power sources. Therefore, we feed back this knowledge into a simulation model of the physical testbed to optimize the channel access for the logistics applications.

### B. Related Work

The PhyNetLab [1] and its underlying platform, the PhyNode [2], have been addressed in numerous research works covering, e.g., the performance availability [4] or the availability of harvested energy from the included photovoltaic cells [5]. In [6] the authors propose energy models for the PhyNode components, which are based on Priced Timed Automata (PTA) and can be used for both use cases, offline energy estimations in simulations, and online energy accounting due to a lightweight integration into the proposed operating system.

Other indoor testbeds for large-scale sensor networks, though not focusing on logistics, are FIT/IoT-Lab [7] and WISEBED [8]. Both networks are public and remotely accessible and provide Software Development Kits (SDKs) to the subscriber. They also include simulation frameworks for offline-testing before deploying firmware into the physical network. However, the nodes in these testbeds have static positions (with the exception of FIT/IoT-Lab recently introducing a hand full of mobile robots), which disqualifies their usage for logistic processes.

A long-term scalability analysis of an outdoor sensor network of 330 nodes is addressed in the *GreenOrbs* project [9]. Here, the authors identified higher link loss rates on nodes which are in proximity to high traffic paths in the network. They suggest, that the Carrier Sense Multiple Access / Collision Avoidance (CSMA/CA) algorithm might not properly detect concurrently transmitting packets and point out the importance of considering the scalability of the radio access scheme during system design.

Collisions on the radio channel reflect a waste of scarce energy, because, in the worst case, none of the colliding messages might be decoded by the receiver and have to be retransmitted. Moreover, this also wastes radio channel resources and computational effort to repeat transmissions. A general survey on energy efficiency in wireless sensor networks is given in [10]. Finally, this motivates us to analyze and optimize the concurrent channel access and reduce packet collisions to a minimum.

The structure of this paper is as follows. In Sec. II we describe the addressed system-scenario and define problem statement in terms of system-specific key performance indicators. Furthermore, we describe in Sec. III the system-in-the-loop architecture at system-level and contrast the simulated model with the PhyNode hardware. After introducing the power consumption model we present the setup for the simulative performance evaluation of the channel access. In Sec. IV we discuss results of a case study and conclude the paper in Sec. V.

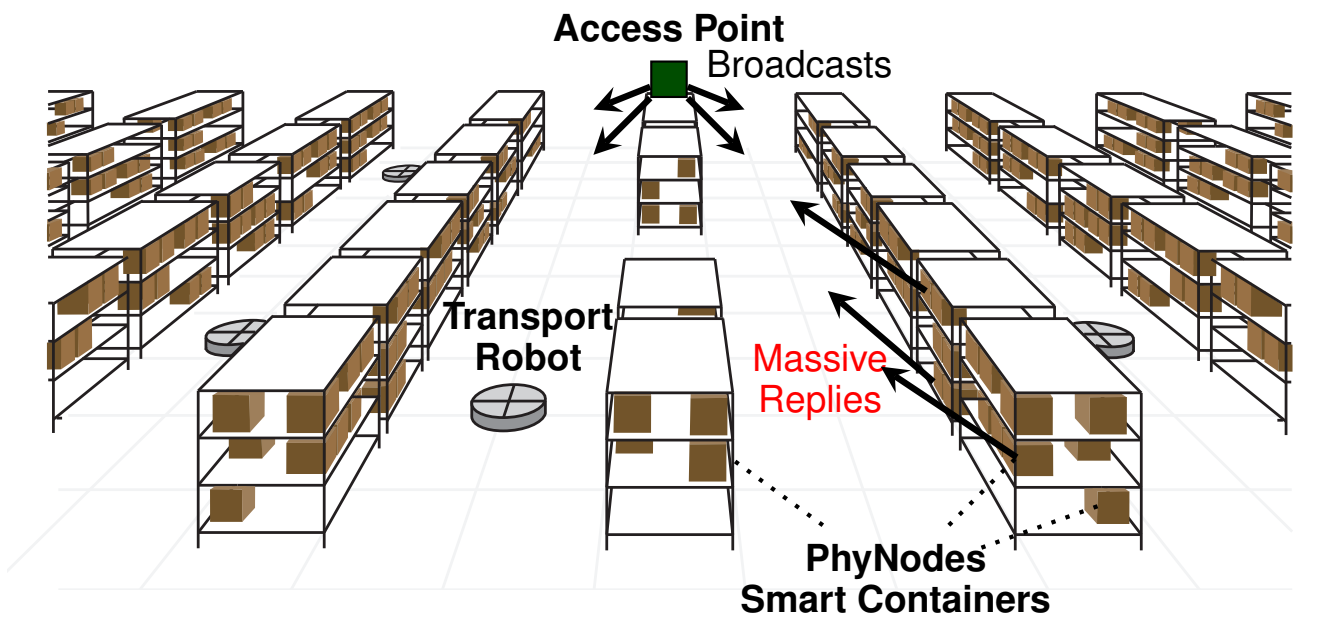

Fig. 2. Deployment and application example of an IoT-based warehouse. Queries for products are broadcast by Access Points (APs) to particularly associated smart containers (PhyNodes). Synchronous replies from matching containers challenge the channel access scheme due a high risk of collisions.

## II. Scenario and Performance Indicators

The System-scenario addresses a large-scale future warehouse, reflected by the PhyNetLab testbed, which stores various products in containers of arbitrary size. While a container is bound to a specific product, products are not limited to one container, but rather are distributed over multiple containers in the warehouse. Each container has an attached PhyNode, with knowledge of the comprised product and the contained quantity. An exemplary overview of such a warehouse deployment is shown in Fig. 2.

The PhyNodes communicate over a radio interface in the SRD band at 868 MHz and are logically attached to the nearest Access Point (AP) in terms of signal strength. The APs are equally distributed in the warehouse and serve as relay stations to the PhyNodes. They collect messages from the PhyNodes and forward them over a local Long Term Evolution (LTE) network and vice versa. This hierarchical network architecture lowers the required transmission power for the PhyNodes and allows a reuse of radio channels in the style of cellular networks. Although APs are battery-powered and occasionally have to be recharged or replaced by robots, they are not scope of this paper. Neighboring APs use different radio channels for communication with the attached PhyNodes to avoid interference. Therefore, it is sufficient to model a single cell covered by one AP to encounter all influencing factors to the behavior and energy consumption of the resource-constrained PhyNodes.

According to this system-scenario, and from the communication's point of view, the most challenging use case of smart warehousing without central stock list, is querying for requested goods. In such case, incoming orders for goods are broadcast into the warehouse in anticipation of sufficient replies to fulfill the order (cf. Fig. 2). The inquirer then selects a subset of replied containers and requests robots to reload the products from the warehouse. Since goods might be distributed over a high number of smart containers, the synchronous replies to such queries challenge the radio channel access in the most intense manner. Without an efficient access procedure, almost all replies to a query would collide on the medium due to similar processing and reply time.

In this paper we focus on the IEEE 802.15.4 CSMA/CA access scheme [11]: Each pending transmission is held back by the transceiver for $b \in \mathbb{N}_0$ Unit Backoff Periods (UBPs). The parameter $b$ is randomly chosen in the interval $\left[0, \ldots, 2^{BE} - 1\right]$ with the backoff exponent $BE$ having the initial value $BE_0 = 3$ in the first attempt. If the following Clear Channel Assessment (CCA) fails due to an occupied channel, $BE$ is incremented by one (with a maximum value of 8) and the backoff procedure is repeated. Otherwise the transmitter sends its packet and resets $BE$ to its initial value for the next transmission. One UBP corresponds to 20 symbol periods and equals to 1 ms in our setup (cf. Tab. I) due to a symbol rate of 20 kSymbols/s and 2-GFSK modulation. Since multiple nodes may still chose equal $b$, this channel access scheme inheres a probability for packet collisions as well as a latency caused by repeatedly failing CCA. Therefore, we define the following performance indicators to quantify the schemes' efficiency.

### A. *System-Specific Key Performance Indicators*

According to the defined warehousing scenario, we specify of the following performance indicators to quantify the efficiency and scalability of the channel access scheme:

**Query Response Ratio (QRR)**: As a consequence of scattering equal products among numerous containers, queries will be replied by multiple containers simultaneously. Due to a residual probability for collisions from the CSMA/CA scheme, only a subset of reply messages will be received by the inquirer. For each query the ratio of collision-free replies in relation to the total number of replying nodes $N$ is defined as $\text{QRR}(N, BE_0) \in [0, 1]$. The QRR shall be maximized during system design, because colliding messages reflect a waste of energy and radio resources. Since a query might be satisfied by a subset of reply messages, we define $\text{QRR}_{\min}$ as the minimum required QRR to ensure proper functionality of the warehouse system.

**Query Response Time (QRT)**: The passed time between sending a product poll into the network and receiving a sufficient fraction ($\geq \text{QRR}_{\min}$) of replies, hence satisfying the query, is defined as $\text{QRT}(\text{QRR}_{\min}, N, BE_0)$. Sine it may happen that a single query does not satisfy the required minimum response ratio $\text{QRR}_{\min}$ due to collisions, the inquirer repeats its query if a message is followed by $T_{\text{wait}}$ seconds of inactivity. The inquirer collects a union set

$$R_{\cup} = \bigcup_k R_k \tag{1}$$

of the previous reply-message sets $R_k$, with $k$ representing the repetition, and checks it against the terminating condition. In these cases QRT reflects the time interval between sending the first query and finally reaching the terminating condition

$$\frac{|R_{\cup}|}{N} \geq \text{QRR}_{\min}. \tag{2}$$

Therefore, QRT reflects the performance of the warehouse system in terms of possible queries per time interval.

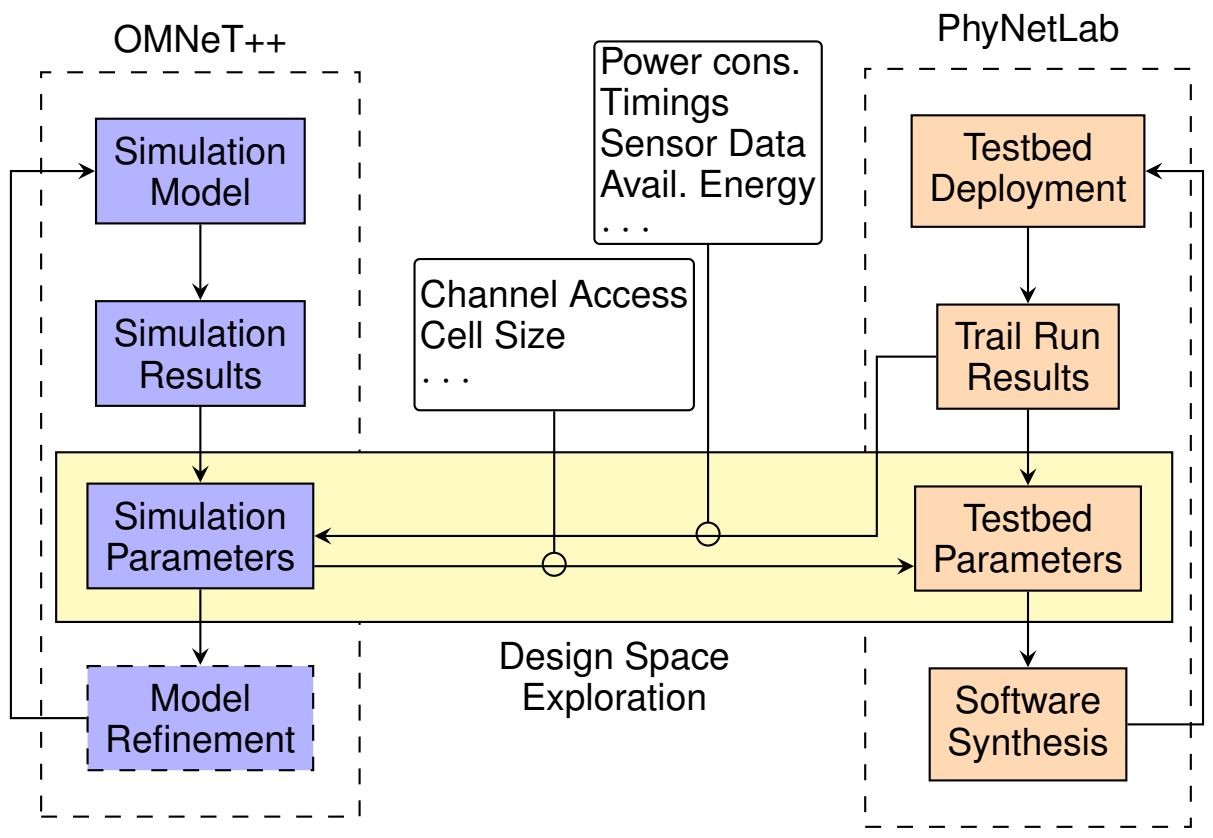

Fig. 3. System architecture for multi-methodical parameter optimization consisting of a real-world, IoT-based warehouse testbed and a simulation model for the radio communication based on OMNeT++ network simulator.

**Energy Consumption per Device**: Each query incurs an energy consumption $E(N, BE_0)$ in each PhyNode for reception, CCA, and transmission of the reply message. In case of repetitive queries to satisfy $\text{QRR}_{\min}$, we later sum up the energy consumption of distinct attempts to compare the overall efficiency of different channel access configurations. Note that even in the case of early satisfying $\text{QRR}_{\min}$, we continue the energy accounting until all nodes finish their transmission, because the energy consumption incurred by those "late" messages still is a consequence of the initial query. Further parameters, e.g., transmission power, packet length, and data rate, are kept constant through the experiments, since their optimization is not in the scope of this paper.

## III. Simulation-based System Model and System Architecture

Motivated by the overlaying challenge of determining the best parametrization of the proposed Internet of Things (IoT)-based warehouse, a multi-methodical approach is used that brings together testbed experiments and simulative evaluations (cf. Fig. 3). While the real-world system provides the application models and the ability to measure the actual properties of the individual hardware components, it cannot be effectively used for parameter studies and large scale evaluations due to the high configuration effort. Instead, those analyses are performed with an OMNeT++-based [12] simulation model. Afterwards, the best parameters are applied into the system software of PhyNetLab and evaluated in trial runs.

OMNeT++ is an Open Source and well-established simulation framework in the communication networks domain. Due to its modular approach, it has been extended by many extension frameworks focusing on specific communication technologies. For the simulative evaluation performed in this paper, the IEEE 802.15.4 model of [13] has been used with the focus on an analysis of the radio channel assessment and its impact on the QRR and the energy consumption of the system components. We modeled the warehouse system including a set of PhyNodes together with an AP working as a relay station. Due to the cellular structure of the network (cf.

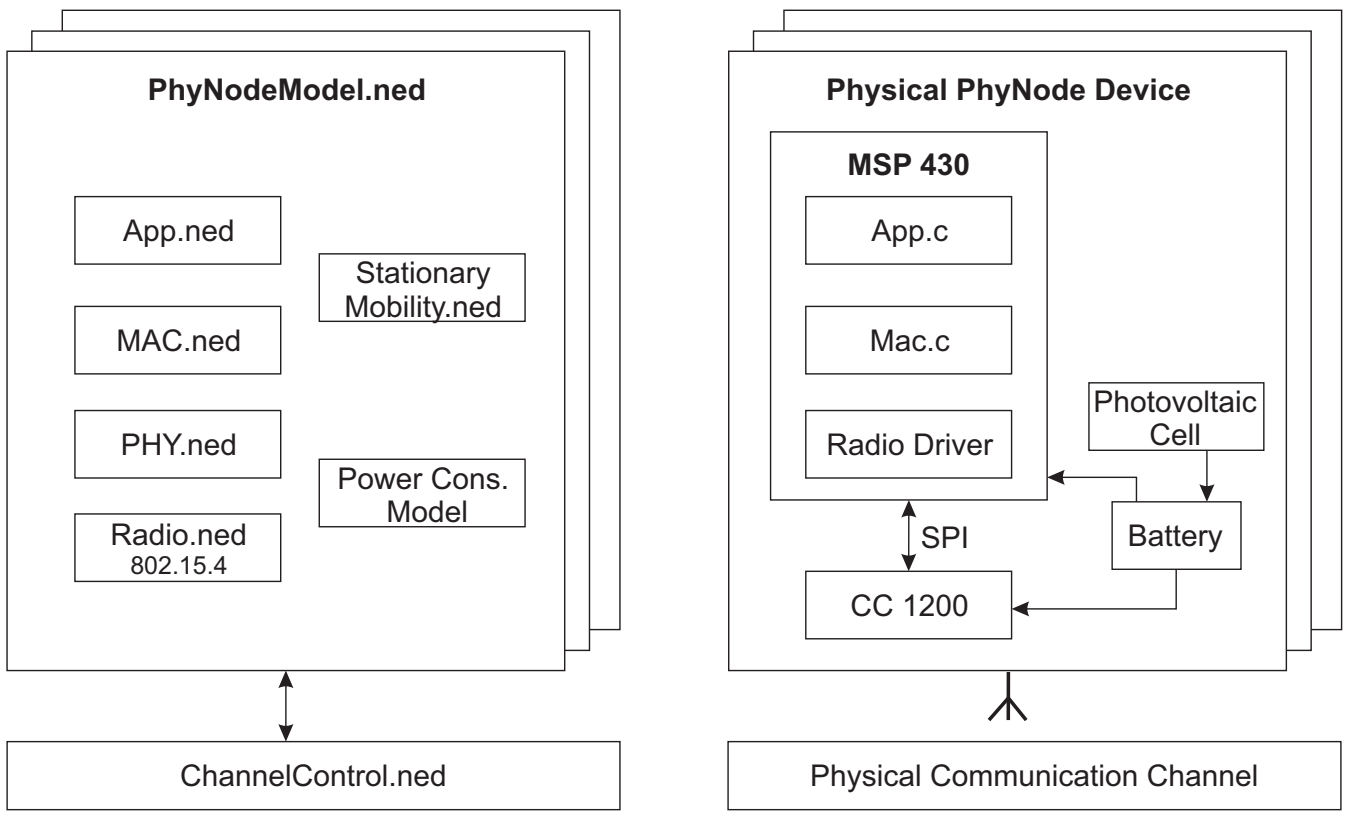

Fig. 4. Architecture of the PhyNode model, and its included 802.15.4 stack components (left) which reflect the physical system platform (right).

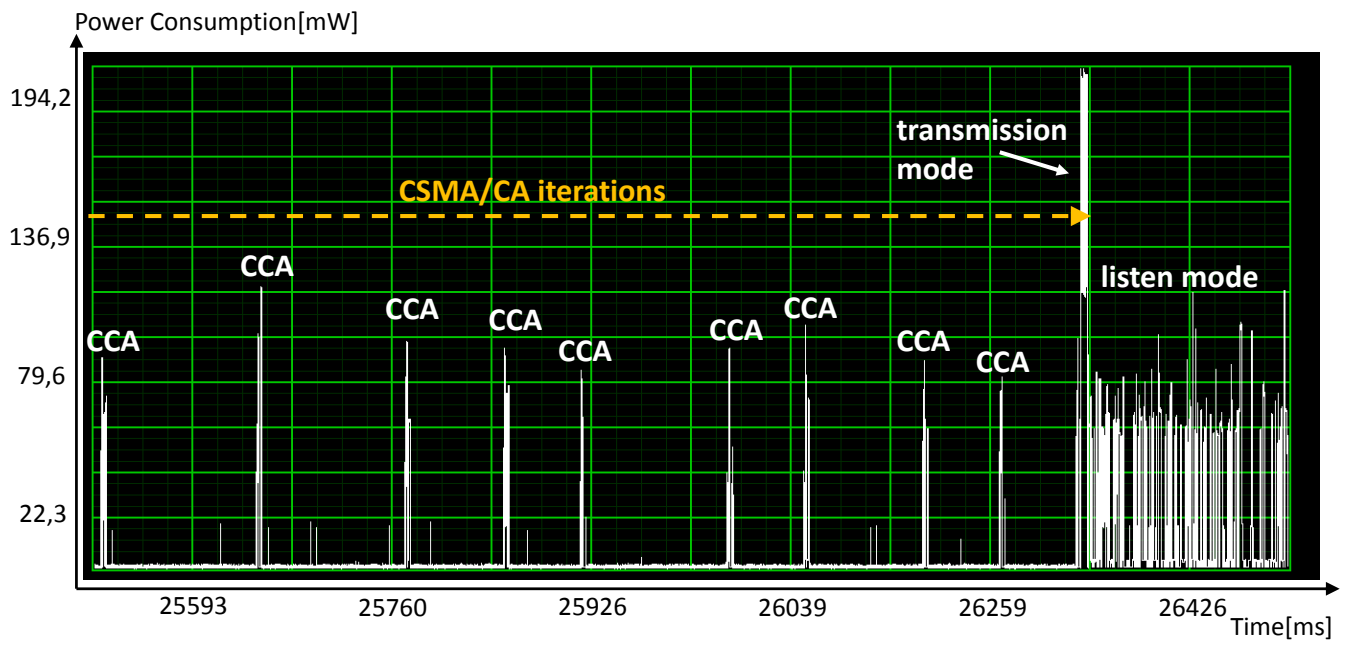

Fig. 5. Measured power consumption of the PhyNode's CC1200 transceiver for replying to a product query in PhyNetLab. Due to concurrency, the CCA fails multiple times before finally transmitting the message.

Sec. II), it is sufficient to simulate a single cell, as neighboring cells use different radio channels for communication with their AP and do not interfere with each other.

The simulation model is composed from numerous units, each reflecting parts of the physical system. Fig. 4 shows a detailed comparison of the simulation architecture and the PhyNode hardware.

Starting from the bottom, the physical radio medium is modeled by the unit `ChannelControl.ned`. It models the signal propagation in terms of path loss, delay and interference due to concurrent medium access. Instances of the radio-interface model `Radio.ned` can subscribe to the channel controller and compute an individual Signal to Interference and Noise Ratio (SINR) estimate for each delivered packet. Based on the receiver's sensitivity and a minimum SINR threshold for a correct reception of a packet, the radio model can mark the received packet as correct or corrupted before passing it to the next layer of abstraction.

While each activity of the physical radio interface (CC1200) dissipates a certain amount of energy from the battery, a power consumption model (cf. next section) keeps track of this dissipation in the simulation, by following state transitions of the radio model.

The remaining layers of the physical system are located in the system software of the PhyNode and are executed by the Micro Controller Unit (MCU). In contrast, the simulation's radio model also includes functionalities of the radio driver, e.g., configuration of transmission power, CCA requests and event handling of incoming transmissions. The actual driver interface is provided by `PHY.ned`.

Address management and the implementation of the CSMA/CA algorithm are part of the Media Access Control (MAC)-layer. It includes the configurable initial backoff exponent $BE_0$, which is a design parameter of the case study in this paper.

The top-layer application provides functionalities of the overlaying logistics process. Since we target an analysis of the channel access procedure, the application currently covers only answering to product queries, which challenges the access scheme in the most intense manner. It will be extended to the full logistics process in future work.

The source code of the simulation model is public available [14] and can be executed on any general purpose computer.

### A. Measurements

To feed the simulation model with power consumption values under realistic conditions, we performed measurements of the physical hardware, especially of the radio transceiver. Since the addressed scenario does not require any sensor data or complex computations, we neglect the power consumption of any peripherals and the low power microcontroller in this scenario. In Fig. 5, we show an exemplary measurement trace of the transceiver's power consumption. We set up a subset of PhyNodes to reply for product queries simultaneously, according to the IEEE 802.15.4 CSMA/CA procedure. In the trace, distinct CCA-attempts appear as short peaks, where the transceiver probes the radio channel for a short interval after each backoff time. Due to a high concurrency on the radio channel, the CCA repeatedly fails nine times. During the highest peak in the trace, the transceiver finally sends its buffered message over the channel. Subsequently, the devices enters the listen mode, and sniffs the radio channel at a high rate to detect other packets on the channel, e.g., new queries.

Based on the traced behavior, the transceiver is modeled as state machine and includes four states `BACKOFF`, `LISTEN`, receive (`RX`), and transmit (`TX`), as shown Fig. 6. Each state is annotated by an average power consumption $P_i$ where $i \in \{\texttt{BACKOFF}, \texttt{LISTEN}, \texttt{RX}, \texttt{TX}\}$, which is obtained from previously described measurements. Although these measurements cover only one specific transceiver, the method is generally applicable on arbitrary devices as well [6].

### B. Setup for the Simulative Performance Evaluation

Based on the proposed model, we performed a case study and set up simulations of up to 410 communicating containers, which are logically attached to one AP located in the center of the arrangement. The setup represents one cell of the entire warehouse. Since neighboring cells do not interfere with each other, the results can be generalized to the full warehouse deployment. The containers are organized in storage racks (cf.

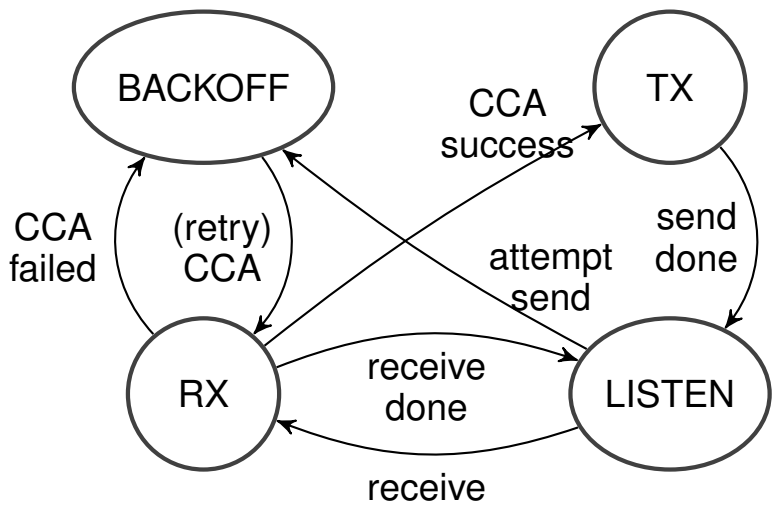

| State | Avg. Power |
|---|---|
| BACKOFF | 1.5 µW |
| LISTEN | 4.5 mW |
| RX | 70 mW |
| TX | 138 mW |

Fig. 6. Power consumption model of the PhyNode's CC1200 transceiver.

Fig. 2) with laterally attached PhyNodes facing the corridor between the shelves.

A single simulation run is initialized by placing $N$ nodes into random locations in the available racks. The nodes represent containers with products of the same kind, hence all nodes will reply to the product query simultaneously. Since nodes with other products do not reply, they are not included explicitly into the setup. After initializing the positions, the AP– representing the inquirer – broadcasts a product query into the network. If he receives at least $\lceil N \cdot \mathrm{QRR}_{\min} \rceil$ different replies, the run terminates after the last node transmits its reply. Otherwise, the AP repeats its query until sufficient different replies reach the AP. After each run, we store the individual energy consumption of the nodes, as well as the QRR and QRT into logfiles.

For a scalability analysis of the CSMA/CA algorithm and its impact on the energy consumption in a large-scale deployment, we subsequently increased $N$ from 10 to 410 nodes and repeated each configuration 20 times. We modified the initial backoff exponent $BE_0$ in the range of 3, 6 and 8, and repeated the simulations for these configurations as well.

Finally, we performed all configurations with $\mathrm{QRR}_{\min}$ set to 80 % and 20 % in order to evaluate the impact of the product distribution in the warehouse to the QRR and energy consumption. An overview of the remaining system parameters, which remained constant, is given in Tab. I.

## IV. Results of Case Study

In this section we present and discuss the results of a first case study, which is accomplished by the proposed system and which we configured as described in the previous section. The results of the average energy consumption per node in Fig. 7 show a continuous increase of energy consumption for growing numbers of concurrently replying nodes. However, the system configuration in terms of initial backoff exponent $BE_0$ and required $\mathrm{QRR}_{\min}$, clearly affects the dissipation of energy in the network. The most upper three curves represent a warehouse deployment with high demands on the reply ratio $\mathrm{QRR}_{\min} = 80\,\%$, which is required if product queries generally demand for a high ratio of the stored goods (of the same kind). In such scenario, setting the initial CSMA/CA backoff exponent $BE_0$ from its default value of 3 to the maximum of 8, leads to energy savings of 49 % for the case of 410 nodes. In case of 150 nodes, this still leads to savings of 44 %, although the benefits of an increased backoff shrink for smaller deployments. While the reader might expect a larger backoff exponent to increase the Query Response Time (QRT), this is generally not the case as Fig. 7 (right) confirms. Instead, a short backoff exponent leads to a high amount of collisions on the radio channel, hence forcing the inquirer to repeat his query multiple times until finally satisfying $\mathrm{QRR}_{\min}$ . A larger backoff reduces the amount of collisions and consequently reduces QRT. In the addressed configuration the QRT shortens by 56 % at 410 replying nodes, and even 63 % at 150 nodes. Therefore, warehouse deployments with high demand on QRR, quickly benefit significantly from the largest backoff configuration, and save both energy and time.

[1]Representing the covered cell by a single AP. The actual warehouse consists of numerous neighboring cells.

TABLE I
System Configuration for Channel Access Analysis

| Parameter | Value |
|---|---|
| **Layout** | |
| Container Dimensions (w, l, h) | 30 cm, 40 cm, and 15 cm |
| Cell Dimensions[1] (w, l, h) | 7.9 m, 8.5 m, and 3.2 m |
| Columns, Rows, Layers | 20, 10, 5 |
| Number of Placed PhyNodes | 10 to 410 |
| **CSMA/CA Configuration** | |
| Repeat Timeout $T_{\text{wait}}$ | 1 s |
| Backoff Exponent $BE_0$ | 3, 6, 8 |
| $\mathrm{QRR}_{\min}$ | 0.8, 0.2 |
| **Radio** | |
| Operating Frequency | 868 MHz |
| Propagation Model | Free Space |
| Transmission Power | 10 dBm |
| Packet Length | 23 B |
| Modulation | 2-GFSK |
| Symbol Rate | 20 kSymbols/s |

In systems with low demands on QRR, e.g., $\mathrm{QRR}_{\min} = 20\,\%$, the impact of $BE_0$ slightly differs from the previous configuration as shown in the lowest three curves in both figures: For high numbers of concurrently replying nodes, $BE_0 = 3$ still incurs the highest energy dissipation in the given setup. However, it undershoots the energy consumption of larger backoff configurations for smaller deployments below 200 nodes. Due to the repeat timeout $T_{\text{wait}}$, QRT for $BE_0 = 3$ still produces the highest delays in this setup.

Comparing the results for $BE_0$ = 6 and 8, we receive overlapping results for QRT, but a slightly lower energy consumption for $BE_0 = 6$. This difference is issued by different degrees of over-satisfying the query in the first and only attempt. The largest $BE_0$ configuration leads to less collisions on the medium and incurs additional packet receptions in those nodes, who early transmitted their own reply message. Although these messages are withdrawn after address check in the MAC layer, they still lead to a slightly increased energy consumption due to transitions in RX state.

Finally the results reveal the limitations of the default 802.15.4 radio channel access scheme for energy-constrained large-scale deployments in case of synchronized replies to broadcast messages. However, the results also show possible solutions by modifying the CSMA/CA configuration and setting the initial backoff exponent to a value, which leads to the lowest energy consumption and/or the shortest query reply time for a given warehouse configuration.

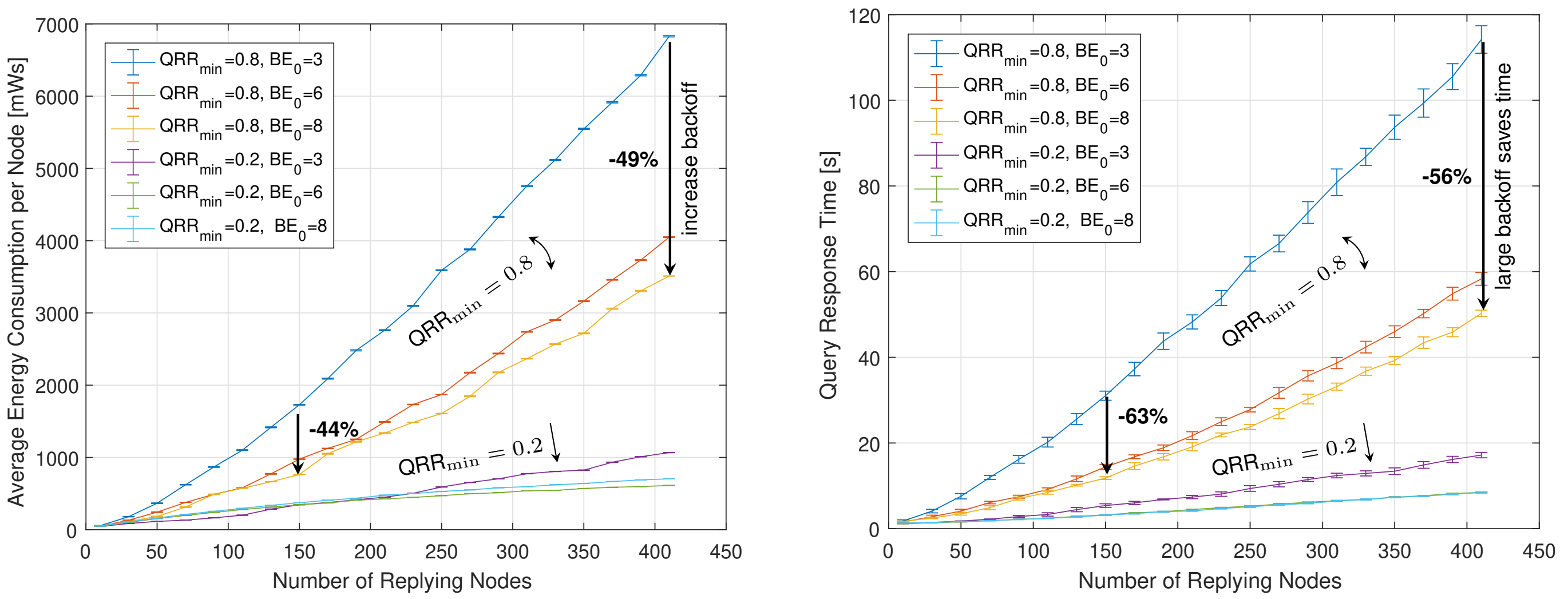

Fig. 7. Energy consumption of PhyNodes (left) and query response time (right) in relation to the number of concurrently replying nodes for different backoff configurations and minimum required query response ratio. The error bars represent confidence intervals.

## V. Conclusion and Future Work

In this paper we have presented a system-in-the loop simulation model of PhyNetLab, a large-scale IoT-based warehouse system. With the objective on energy-efficient communication, we focused on the channel access procedure and analyzed the performance in the highly competitive use case of querying for distinct products in the warehouse.

We showed, that the default 802.15.4 CSMA/CA procedure leads to high amounts of collisions on the radio channel which incur a long query delay and a high energy consumption in the replying nodes. However by adapting the initial backoff exponent of the CSMA/CA algorithm, we showed that the energy dissipation is reduced in the order of 50 % for large-scale deployments, in which 410 nodes simultaneously reply to to queries for a single product. Additionally, this approach also reduces the Query Response Time (QRT) in the same order of magnitude, since queries are fulfilled by less repetitions.

In future work we will merge our system with a simulation model of the actual logistics processes, which will introduce mobility and inter-cell dynamics, e.g., handover between cells. Furthermore, we include multi-channel approaches like frequency division duplex (FDD) and evaluate their suitability for such systems.

## Acknowledgment

Part of the work on this paper has been supported by Deutsche Forschungsgemeinschaft (DFG) within the Collaborative Research Center SFB 876 "Providing Information by Resource-Constrained Analysis", project A4.

## References

[1] A. K. Ramachandran Venkatapathy, M. Roidl, A. Riesner, J. Emmerich, and M. ten Hompel, "PhyNetLab: Architecture design of ultra-low power wireless sensor network testbed," in *IEEE 16th International Symposium on A World of Wireless, Mobile and Multimedia Networks*. IEEE, 2015, pp. 1–6.

[2] A. K. Ramachandran Venkatapathy, A. Riesner, M. Roidl, J. Emmerich, and M. ten Hompel, "PhyNode : An intelligent, cyber-physical system with energy neutral operation for PhyNetLab," in *Proceedings of Smart SysTech; European Conference on Smart Objects, Systems and Technologies;*. VDE-Verl, 2015, pp. 1–8.

[3] R. Falkenberg, M. Masoudinejad, M. Buschhoff, A. K. Ramachandran Venkatapathy, D. Friesel, M. ten Hompel, O. Spinczyk, and C. Wietfeld, "PhyNetLab: An IoT-based warehouse testbed," in *2017 Federated Conference on Computer Science and Information Systems (FedCSIS)*, Prague, Czech Republic, sep 2017, pp. 1051–1055.

[4] M. Roidl, J. Emmerich, A. Riesner, M. Masoudinejad, D. Kaulbars, C. Ide, C. Wietfeld, and M. T. Hompel, "Performance availability evaluation of smart devices in materials handling systems," in *2014 IEEE/CIC International Conference on Communications in China - Workshops (CIC/ICCC)*, 2014.

[5] M. Masoudinejad, M. Kamat, J. Emmerich, M. ten Hompel, and S. Sardesai, "A gray box modeling of a photovoltaic cell under low illumination in materials handling application," in *3rd International Renewable and Sustainable Energy Conference (IRSEC)*, 2015, pp. 1–6.

[6] M. Buschhoff, R. Falkenberg, and O. Spinczyk, "Energy-aware device drivers for embedded operating systems," *SIGBED Rev.*, 2018.

[7] FIT consortium, "FIT/IoT-LAB Very large scale open wireless sensor network testbed." [Online]. Available: https://www.iot-lab.info/

[8] G. Coulson, B. Porter, I. Chatzigiannakis, C. Koninis, S. Fischer, D. Pfisterer, D. Bimschas, T. Braun, P. Hurni, M. Anwander, G. Wagenknecht, S. P. Fekete, A. Kröller, and T. Baumgartner, "Flexible experimentation in wireless sensor networks," *Commun. ACM*, vol. 55, no. 1, pp. 82–90, Jan. 2012.

[9] Y. Liu, Y. He, M. Li, J. Wang, K. Liu, and X. Li, "Does wireless sensor network scale? a measurement study on greenorbs," *IEEE Transactions on Parallel and Distributed Systems*, 2013.

[10] T. Rault, A. Bouabdallah, and Y. Challal, "Energy efficiency in wireless sensor networks: A top-down survey," *Computer Networks*, vol. 67, no. Supplement C, pp. 104 – 122, 2014.

[11] "IEEE standard for low-rate wireless networks," *IEEE Std 802.15.4-2015 (Revision of IEEE Std 802.15.4-2011)*, pp. 1–709, April 2016.

[12] A. Varga and R. Hornig, "An overview of the OMNeT++ simulation environment," in *Proceedings of the 1st International Conference on Simulation Tools and Techniques for Communications, Networks and Systems & Workshops*, ser. Simutools '08. Brussels, Belgium: ICST, 2008, pp. 60:1–60:10.

[13] M. Kirsche and M. Schnurbusch, "A new IEEE 802.15.4 simulation model for OMNET++ / INET," in *Proc. of 1st OMNeT++ Community Summit*, 2014.

[14] R. Falkenberg, "falkenber9/phynetlabsim: Initial release," Nov. 2017. [Online]. Available: https://doi.org/10.5281/zenodo.1054424